\documentclass[journal,twoside,web]{ieeecolor}
\usepackage{generic}
\usepackage{cite}
\usepackage{amsmath,amssymb,amsfonts}
\usepackage{graphicx}
\usepackage{textcomp}
\usepackage{booktabs}
\usepackage{array}
\usepackage{multirow}
\usepackage{tabularx}
\usepackage{makecell}
\usepackage{hyperref}
\hypersetup{hidelinks}
\newcolumntype{C}{>{\centering\arraybackslash}X}

\def\BibTeX{{\rm B\kern-.05em{\sc i\kern-.025em b}\kern-.08em
    T\kern-.1667em\lower.7ex\hbox{E}\kern-.125emX}}
\title{Cross-View Attention Fusion Net: A Prior-Guided Dual-View Representation Learning for Cardiac Output Estimation in Short-Term Photoplethysmography Signals}
\author{Yaowen Zhang, Bo Cui, Libera Fresiello, Peter H. Veltink, \IEEEmembership{Senior Member, IEEE}, Dirk W. Donker, and Ying Wang
\thanks{This work was supported in part by the Scholarship from China Scholarship Council (CSC). (Corresponding author: Ying Wang.)}
\thanks{This work involved human subjects or animals in its research. Approval of all ethical and experimental procedures and protocols was granted by the Ethics Committee of Computer and Information Science, University of Twente, under Approval No. 240831.}
\thanks{Yaowen Zhang, Bo Cui, Peter H. Veltink and Ying Wang are with the Department of Biomedical Signals and Systems, University of Twente, 7522 NB Enschede, The Netherlands (e-mail: y.zhang-12@utwente.nl; p.h.veltink@utwente.nl; imwywk@gmail.com).}
\thanks{Libera Fresiello and Dirk W. Donker are with the group of Cardiovascular and Respiratory Physiology, University of Twente, 7522 NB Enschede, The Netherlands (e-mail: l.fresiello@utwente.nl; d.w.donker@utwente.nl).}}

\begin{document}
\maketitle

\begin{abstract}
Cardiac output (CO) estimation from photoplethysmography (PPG) could support noninvasive wearable hemodynamic monitoring. However, feature-engineered approaches depend on reliable pulse delineation, while raw-signal deep learning often underuses established PPG transformations. We present the Cross-View Attention Fusion Network (CVAF-Net), a prior-guided dual-view architecture that combines signal knowledge with data-driven learning. Separate encoders process raw PPG and an eight-channel feature sequence map (FSM), and cross-view attention integrates their complementary representations without explicit pulse-cycle delineation. We evaluated CVAF-Net using subject-wise five-fold cross-validation on 5-, 15-, and 30-second segments from simulated pulse waves (3,323 subjects), vasoconstriction provocation (79 subjects), and resting/cycling activities (10 subjects). CVAF-Net achieved minimum mean absolute error (MAE) values of 0.19 L/min on simulated data (mean absolute percentage error, 3.95\%) and 1.20 L/min on measured data, with performance comparable to the Transformer baseline. Removing either encoder or replacing cross-view attention with concatenation degraded performance. Predicted CO showed inverse population-level associations with age (Spearman’s $\rho$ = -0.367 to -0.274) and subject-level associations with heart rate (median $\rho$ up to 0.894) and systemic vascular resistance (median $\rho$ down to -0.740) in the resting/cycling dataset. Synthetic motion-artifact testing assessed sensitivity to signal degradation. Moreover, CVAF-Net required over twelvefold fewer floating-point operations than the Transformer. These findings support prior-guided dual-view learning as an accurate, computationally efficient, physiologically consistent, and comparatively robust approach for short-window PPG-based CO estimation, while external clinical validation remains necessary before deployment.
\end{abstract}


\section{Introduction}
Cardiovascular diseases (CVDs), notably chronic heart failure (CHF), remain the leading causes of global morbidity, requiring proactive management to improve prognosis and prevent acute decompensation \cite{WHO2025,Roth2020}. As a primary indicator of cardiac performance, cardiac output (CO) provides critical insights into hemodynamic status, with declining values often preceding clinical symptoms of circulatory shock \cite{Rusinaru2021,Vahdatpour2019}. Consequently, enabling accurate, non-invasive CO monitoring in home settings is essential for the continuous management of chronic CVDs and timely clinical intervention.

While thermodilution remains the clinical gold standard for CO measurement, its invasiveness limits its use to intensive care \cite{Vahdatpour2019}. Current non-invasive alternatives, such as echocardiography or bioimpedance, often face challenges regarding operator dependency or signal interference \cite{Geerts2011, Grensemann2018, Saugel2018}. Owing to its low cost and widespread integration into wearables, photoplethysmography (PPG) has emerged as an ideal modality for daily CO monitoring, leveraging its capacity to capture peripheral blood volume dynamics that fundamentally reflect stroke volume and vascular resistance \cite{Charlton2022}.

Current approaches to PPG-based CO estimation broadly fall into two categories: feature-engineered machine learning (ML) and end-to-end deep learning (DL) methods. Conventional ML frameworks rely heavily on morphological and dynamical PPG descriptors, such as pulse amplitude, systolic rise time, and entropy-based metrics \cite{Lee2013, Wang2009}. However, these features depend strictly on accurate pulse-cycle segmentation and fiducial-point detection. Errors in identifying pulse onsets or extrema can directly bias the extracted descriptors, thereby degrading CO estimation. Furthermore, reducing the continuous waveform into predefined scalar features permanently discards critical temporal information that the regression model cannot recover. This limitation is especially problematic in ambulatory and exercise scenarios, where motion artifacts, baseline wander, low perfusion, and irregular waveforms routinely disrupt reliable beat delineation. Therefore, fixed-duration segment-based modeling provides a compelling alternative by minimizing the need for explicit pulse segmentation.

While end-to-end DL models avoid these preprocessing requirements by learning directly from raw PPG segments, they introduce a different representational bottleneck \cite{Ipar2025, G2025}. The physiological information related to CO is distributed across local morphology, temporal dynamics, oscillatory patterns, and multiscale signal characteristics. Constraining the model to a single, one-dimensional raw input stream forces one learning pathway to discover these complex structures simultaneously. Processing finite and potentially noisy data without explicit guidance--ignoring the signal characteristics already known to be informative in PPG analysis--limits the model's ability to fully capture the underlying physiological dynamics.

Signal transformations offer an effective way to introduce this structural guidance without compressing the waveform into handcrafted scalar features. For example, derivatives highlight slope and acceleration; frequency-domain and Hilbert representations reveal oscillatory, envelope, and phase structures; and wavelet decompositions capture multi-scale temporal information. Rather than acting as direct physiological measurements, these transformations serve as complementary views of the original PPG signal. The central methodological challenge, therefore, is to preserve the complete raw temporal waveform while simultaneously integrating these prior-guided representations, enabling the model to learn how these different views interact for robust CO estimation.

Motivated by this gap, this paper presents the Cross-View Attention Fusion Network (CVAF-Net), a prior-guided dual-view deep learning framework for CO estimation from short fixed-duration PPG segments. Here, "dual-view" denotes the deliberate separation of the same PPG segment into two complementary inputs: a raw temporal view and a feature sequence map (FSM) view derived from established signal transformations. We hypothesize that CO-related information can be better captured when the model learns both views separately and then fuses them through attention mechanism, rather than treating all inputs as a single flat vector or a single concatenated stream. The methodological highlights are summarized as follows:
\begin{enumerate}
    \item We propose CVAF-Net, a prior-guided dual-view architecture that separately encodes raw PPG and an eight-channel FSM and integrates them through cross-view attention. The FSM-derived representation provides structural context for selecting complementary information from the raw temporal representation, enabling learned interaction between prior-guided and data-driven signal representations within a unified framework.
    \item We formulate CO estimation as a short fixed-duration segment task using 5-second, 15-second, and 30-second PPG windows, reducing dependence on accurate beat detection and supporting rolling-window wearable implementation in real world.
    \item We evaluate the model on accuracy and physiological plausibility aspects across varying short-term lengths and multiple diverse datasets to ensure robust performance and generalization. The datasets include: i) standardized in-silico datasets; ii) datasets collected during high motion variability physiological states (exercise); iii) datasets involving significant hemodynamic changes across different age groups (vasoconstriction provocation studies).
\end{enumerate}

\section{Method}
\begin{figure*}[htbp]
\centerline{\includegraphics[width=0.98\textwidth]{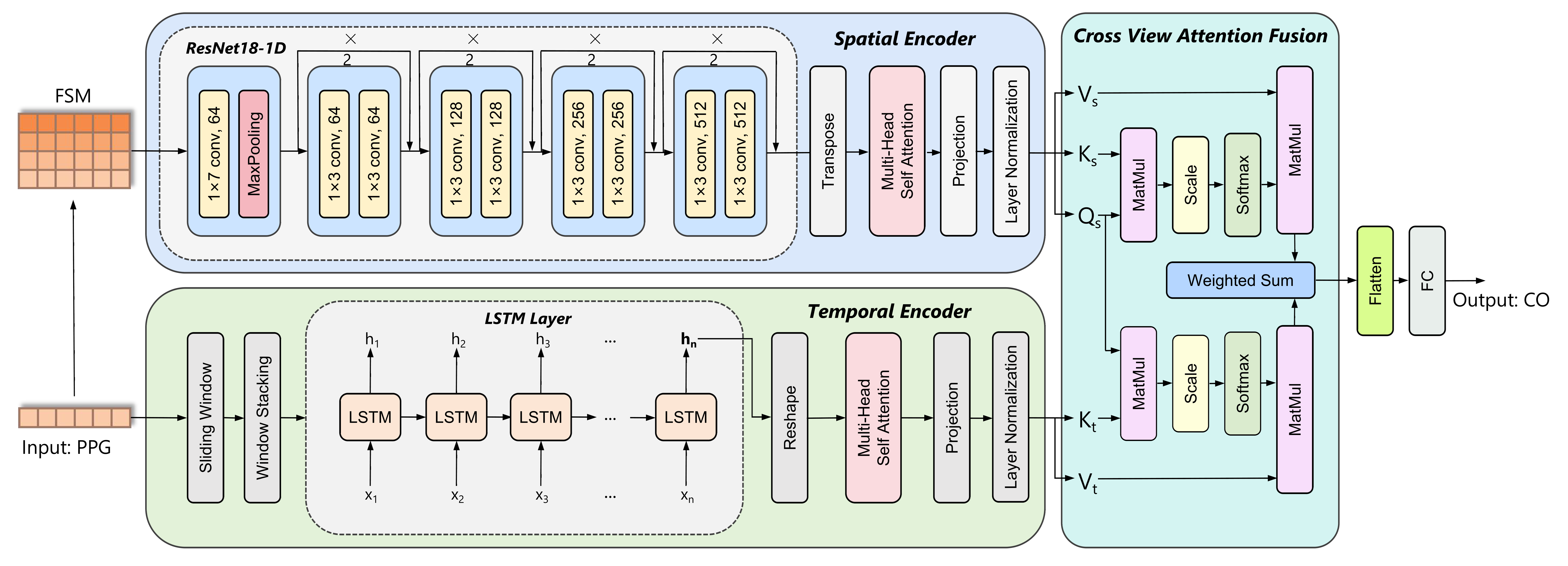}}
\caption{Schematic diagram of the CVAF-Net architecture. The framework consists of three modules: a temporal encoder, a spatial encoder and a cross view attention fusion module. FSM: feature sequence map; $Q_s$, $K_s$, $V_s$: query, key, value from spatial encoder; $K_t$, $V_t$: key, value from temporal encoder; FC: fully-connected layer; CO: cardiac output.}
\label{fig1}
\end{figure*}

The architecture of cross view attention fusion network (CVAF-Net) is illustrated in Figure \ref{fig1}. The network comprises three main components: a temporal encoder, a spatial encoder, and a cross view attention fusion module. The temporal encoder learns latent dynamics from the raw PPG segment, the spatial encoder learns structured patterns from the prior-guided FSM, and the fusion module learns how the two views interact before estimating CO. These are followed by a prediction head acting as the decoder, which flattens the fused representations and maps them through fully connected layers to yield the final estimated CO.

\subsection{Temporal encoder}
\label{TE}
The temporal encoder processes the PPG time-series data. It is designed to capture the dynamic and sequential dependencies inherent in the cardiac cycle. As shown in Figure \ref{fig1}, this is achieved using a sliding window long short term memory (LSTM) augmented with a multi-head self-attention mechanism to focus on the most salient temporal segments.

First, the input PPG signal $x \in \mathbb{R}^{B \times L}$ is processed by a sliding-window LSTM, using a window size of 64 samples and a stride of 32 samples to generate overlapping local contexts from each fixed-duration segment; the resulting number of windows is denoted as $N_t$, the batch size as $B_t$, and the sequence length as $L$. Each windowed segment is encoded by an LSTM unit with a hidden dimensionality $h=64$.

Consequently, the sliding window LSTM block generates an intermediate representation $H$:

\begin{equation}
\label{eq:sliding_window}
H = \text{SlidingWindowLSTM}(x) \in \mathbb{R}^{B_t \times N_t \times h}
\end{equation} 

To integrate dependencies across different windows, the hidden vectors $H$ are passed into a multi-head self attention (MSA) block. After transposing the sequence and hidden dimensions, the temporal attended feature $A_t$ is computed as:
\begin{equation}
\label{eq:MSA_temp}
\begin{aligned}
A_t &= \text{MSA}(H) \\
&= \text{Concat}(head^t_1, \dots, head^t_n)W_t^O
\end{aligned}
\end{equation}
where $\text{Concat}(\cdot)$ denotes the concatenation operation that merges the representations from $n$ independent attention heads into a single vector. $n=8$ is the number of attention heads. $W_t^O \in \mathbb{R}^{h \times h}$ is a learnable temporal output projection matrix that aggregates the concatenated information from all heads back into the hidden dimensionality $h$. Each individual head $head^t_i$ within the $n$-head MSA layer is defined as:
\begin{equation}
\label{eq:Attention_temp}
head^t_i = \text{Softmax}\left(\frac{Q^t_i (K^t_i)^T}{\sqrt{d_k}}\right)V^t_i
\end{equation}
where $\text{Softmax}(\cdot)$ is the activation function applied row-wise to normalize the attention scores into a probability distribution, ensuring the sum of weights for each window equals one. $Q^t_i, K^t_i, V^t_i \in \mathbb{R}^{N \times d_k}$ represent the Query, Key, and Value matrices for the $i$-th head, respectively. These are obtained through linear projections $Q^t_i = HW_{i,t}^Q$, $K^t_i = HW_{i,t}^K$, and $V^t_i = HW_{i,t}^V$, where $W^Q_{i,t}, W^K_{i,t}, W^V_{i,t} \in \mathbb{R}^{N \times d_k}$ are learned weight parameters. $d_k = h/n$ is the dimensionality of each attention head, which acts as a scaling factor. $\sqrt{d_k}$ is the scaling term used to prevent the dot products from growing too large in magnitude, which would otherwise push the Softmax function into regions with extremely small gradients.

Finally, the attended features $A_t$ are mapped to the embedding space $d_t = 128$ and refined through layer normalization to ensure training stability and facilitate effective cross-modal interaction:
\begin{equation}
\label{eq:LayerNorm_temp}
T = \text{LayerNorm}(A_t W_{pt} + b_{pt}) \in \mathbb{R}^{B_t \times N_t \times d_t}
\end{equation}
where $W_{pt} \in \mathbb{R}^{h \times d}$ and $b_{pt} \in \mathbb{R}^d$ are learnable projection weights and bias parameters, respectively. The resulting tensor $T$ constitutes the temporal context for subsequent fusion.

\subsection{Spatial encoder}
\label{SE}
The spatial encoder operates on a feature map derived from the PPG signal. This map can be constructed from handcrafted features or through transformations, representing the signal in a structured, image-like format. As depicted in Figure \ref{fig1}, a one-dimensional ResNet-18 (ResNet18-1D), followed by a multi-head self-attention layer, is employed to extract high-level spatial and structural features from this map.

We utilized a multi-channel feature sequence map (FSM) comprising eight key features extracted from the PPG signal: first derivative, second derivative, Fast Fourier transform (FFT) full amplitude, Hilbert transform, discrete stable wavelet transform (DSWT) approximation coefficients (a) as well as first, second, and third detail coefficients (d1, d2, d3). The length of each feature equals the length of the PPG signal segment. The first derivative and second derivative features are included to highlight the rate of change and acceleration of the signal, respectively, helping to identify rapid variations in blood flow and cardiac activity \cite{Elgendi2012}. The FFT full amplitude feature provides frequency-domain information which reflects the periodic characteristics of the PPG signal \cite{MejiaMejia2022}. The Hilbert transform is employed to extract the instantaneous envelope and phase, capturing the dynamic signal intensity and frequency modulation patterns over time \cite{Huang1998}. Additionally, we extract DSWT-db2 approximation and detail coefficients a, d1, d2, d3. The mother wavelet db2 was chosen based on its similarity with the single beat PPG morphology \cite{John2022}. These coefficients enable multi-resolution analysis, capturing both low-frequency trends and high-frequency fluctuations in the signal. 

The combination of these features offers a comprehensive representation of the PPG signal, integrating time-domain, frequency-domain, and multi-scale information. By concatenating these eight features into a multi-channel sequence map, we provide spatial encoder with a rich, diverse input thus allow it to learn complex patterns in the PPG data with the guidance of prior human knowledge of PPG feature engineering.

The Spatial Encoder extracts high-level features from the FSM $X \in \mathbb{R}^{B_s \times C \times T_s}$ using ResNet18-1D, where $B_s = 32$ denotes the batch size, $C=8$ denotes the number of input feature channels, and $T_s$ is the length of the signal segment. The FSM-encoded feature map, which represents the signal's multi-scale structural information, is obtained as:
\begin{equation}
\label{eq:resnet_out}
R = \text{ResNet18-1D}(X) \in \mathbb{R}^{B_s \times C_r \times N_s}
\end{equation}
where $C_r = 512$ is the expanded feature dimensionality at the output of the final residual block, and $N_s$ is the reduced temporal resolution resulting from the strided convolutions and pooling operations within the ResNet architecture. $R$ is transposed as $R_{seq} = R^T \in \mathbb{R}^{B_s \times N_s \times C_r}$ and processed by a spatial MSA block to reweight feature representations along the reduced temporal dimension:
\begin{equation}
\label{eq:MSA_spatial}
\begin{aligned}
A_s &= \text{MSA}(R_{seq}) \\
&= \text{Concat}(head^s_1, \dots, head^s_n)W_s^O
\end{aligned}
\end{equation}
where $W_s^O \in \mathbb{R}^{C_r \times C_r}$ is the learned output projection matrix for the spatial branch and $n=8$ is the number of attention heads. The individual structural attention heads are computed as:
\begin{equation}
\label{eq:Attention_spatial}
head^s_i = \text{Softmax}\left(\frac{Q^s_i (K^s_i)^T}{\sqrt{d_s}}\right)V^s_i
\end{equation}
where $Q^s_i, K^s_i, V^s_i \in \mathbb{R}^{N_s \times d_s}$ are the Query, Key, and Value matrices projected from $R_{seq}$ using branch-specific attention parameters $W_{i,s}^Q, W_{i,s}^K, W_{i,s}^V \in \mathbb{R}^{C_r \times d_s}$. $d_s = C_r/n$ is the dimensionality of each head, and $\sqrt{d_s}$ serves as the scaling factor to stabilize training gradients.
Finally, the attended FSM representations $A_s$ are projected into a joint embedding space and normalized to facilitate fusion with the temporal branch:
\begin{equation}
\label{eq:LayerNorm_spatial}
S = \text{LayerNorm}(A_s W_{ps} + b_{ps}) \in \mathbb{R}^{B_s \times N_s \times d_s}
\end{equation}
where $W_{ps} \in \mathbb{R}^{C_r \times d_s}$ and $b_{ps} \in \mathbb{R}^d_s$ are learned projection parameters that map the structural features to the unified fusion dimension $d_s = 128$. The tensor $S$ represents the final structural modality representation derived from the FSM.

\subsection{Cross view attention fusion}
The feature sequences generated by the Temporal and Spatial encoders are treated as two distinct views of the PPG data. These sequences are then fed into a cross view attention fusion, as shown in Figure \ref{fig1}, which performs deep, attention-based fusion to model the complex interdependencies between the two views before making the final CO prediction. To integrate the temporal features $T$ and FSM-based features $S$ obtained from the aforementioned encoders, we propose a Cross View Attention Fusion (CVAF) module. We use the FSM-based representation as the query because this branch encodes prior-guided structural information from PPG signal transformations; the query then selects temporal information from the raw PPG branch that is relevant under that structural context. This design operationalizes the study vision of combining human prior knowledge with data-driven feature learning rather than relying on either source alone. First, features are projected into a joint space:
\begin{equation}
\begin{aligned}
Q_s &= S W_Q^s, & K_s &= S W_K^s, & V_s &= S W_V^s,\\
K_t &= T W_K^t, & V_t &= T W_V^t. &&
\end{aligned}
\end{equation}
where $W_Q^s, W_K^s, W_V \in \mathbb{R}^{d \times d}$ are learnable fusion matrices. Using the spatial query $Q_s$, two distinct attention maps are generated:
\begin{equation}
\begin{aligned}
\alpha_{\mathrm{intra}}
  &= \operatorname{Softmax}\!\left(\frac{Q_s K_s^T}{\sqrt{d}}\right),\\
\alpha_{\mathrm{cross}}
  &= \operatorname{Softmax}\!\left(\frac{Q_s K_t^T}{\sqrt{d}}\right).
\end{aligned}
\end{equation}
where $\alpha_{intra}$ captures spatial self-relations and $\alpha_{cross}$ captures cross-modal dependencies. The attended values are aggregated to form the fused representation $V_f$ using learnable scalars $c_1$ and $c_2$:
\begin{equation}
\begin{aligned}
V_f ={}& c_1 (\alpha_{\mathrm{intra}} V_s)
       + c_2 (\alpha_{\mathrm{cross}} V_t)\\
     &\in \mathbb{R}^{B \times N \times d}.
\end{aligned}
\end{equation}
where $c_1$ and $c_2$ are learnable coefficients, initialized to 1.0, that allow the network to adaptively balance the influence of the structural view and the temporal view during training. The fused representation $V_f$ is first compressed into a segment-level feature vector $V_{flat}$ via global average pooling across the $N$ windows:
\begin{equation}
V_{flat} = \frac{1}{N} \sum_{i=1}^{N} V_f[:, i, :] \in \mathbb{R}^{B \times d}
\end{equation}

Finally, $V_{flat}$ is passed through a multi-layer perceptron (MLP) to compute the CO value:
\begin{equation}
\begin{aligned}
F ={}& \operatorname{Dropout}\!\left(
       \operatorname{ReLU}(V_{\mathrm{flat}} W_{fc} + b_{fc})\right)\\
     &\in \mathbb{R}^{B \times d}.
\end{aligned}
\end{equation}
where $W_{fc} \in \mathbb{R}^{d \times d_{out}}$ and $b_{fc}$ are the weights and biases of the fully connected layers. $\text{Dropout}$ is applied with a rate of 0.1 to prevent overfitting. The output CO represents the non-invasively estimated cardiac output in L/min, ensuring that the latent physiological patterns learned by CVAF-Net are effectively translated into clinically relevant metrics.

\section{Experiments}
To evaluate the estimation performance of the proposed CVAF-Net, we extensively conducted estimation tasks on three datasets. The datasets introduction, data pre-processing, experimental details, results and analysis are provided in this section.

\subsection{Datasets and pre-processing}
Three datasets, including two public datasets and one dataset collected in our previous studies \cite{YZhang2025EMBC,Zhang2026}, were employed to train and validate CVAF-Net, each serving a distinct purpose in evaluating its performance. The synthetic nature of the Simulated Pulse Wave dataset \cite{Charlton2019} allows for precise labeling and exploration of the model's generalization across a wide range of physiological parameters. The Vasoconstriction Provocation dataset \cite{Mol2020} provides extensive data for evaluating performance under real-life vascular changes across different age groups. The Resting and Cycling dataset \cite{Yaowendata2026} is used to assess performance under varying physical conditions and dynamic movement.

\subsubsection{Simulated Pulse Wave dataset}
The Simulated Pulse Wave (SPW) in-silico database comprises 4,374 virtual healthy subjects aged 25-75 years, generated using a one-dimensional (1-D) computational model of arterial pulse wave propagation with Windkessel boundary conditions. For each virtual subject, cardiac, vascular, and arterial blood properties were varied within physiological ranges and validated against literature-reported values and morphologies \cite{Charlton2022}. In this study, we used the noiseless simulated finger PPG waveforms, sampled at 500 Hz and normalized to (0,1). Each recording contains a single PPG beat and is paired with the corresponding simulated cardiac output (CO, L/min).

To simulate continuous recordings collected by wearable devices, we first downsampled each PPG simulated signal from the original 500 Hz to 64 Hz to match typical sampling rates used in wearable devices. The simulated CO value corresponding to each original PPG simulated signal was regarded as reference value for the generated segment. To ensure physiological plausibility, segments with CO less than 3.6 L/min were excluded from the analysis according to the observations in CO assessment via pulsed-wave Doppler echocardiography in 3090 healthy subjects, which study reported a 5-95th percentile of 3.6 to 6.9 L/min for all genders \cite{Rusinaru2021}. 

\subsubsection{Vasoconstriction Provocation dataset}
Vasoconstriction Provocation (VP) dataset \cite{Mol2020} was collected from a cohort of 86 subjects, including 55 younger adults (age $< 65$ years) and 31 older adults (age $> 70$ years), during various vasoconstriction provocation protocols, such as the cold pressor test, active stand, and head-up tilting. The data we applied was obtained through 1000 Hz sampled finger PPG from custom made sensor, and beat-to-beat cardiac output estimated by the Finapres device using the Modelflow® pulse-contour analysis of the continuous finger arterial blood pressure waveform (ABPM). Each recording lasts around 35 minutes. Raw PPG was filtered via a third order Butterworth bandpass filter, normalized to (0, 1) and synchronized to beat-to-beat CO on the same time scale.

For model evaluation, VP PPG recordings were downsampled from 1000 Hz to 64 Hz. The median CO value within each segment was used as the reference label. The lower exclusion threshold of 3.6 L/min followed the 5th percentile of healthy adult CO reported \cite{Rusinaru2021}. The upper exclusion threshold was set to 11.1 L/min by applying the 61\% maximum CO increase observed during active standing \cite{Tanaka1996} to the 95th percentile resting CO value of 6.9 L/min reported \cite{Rusinaru2021}. ($6.9 \cdot 1.61 = 11.1$ L/min). Segments outside this range were removed after segmentation to exclude physiologically implausible segment-level labels.

\subsubsection{Resting and Cycling dataset}
The dataset, named Resting and Cycling (RC) dataset \cite{Yaowendata2026}, which applied in our previous works \cite{Zhang2026, YZhang2025EMBC}, was collected in our own experiment including 10 healthy subjects aged 20-30 through two around 30-minute trials. In each trial, subjects were asked to conduct activities including 5 minutes' sitting, 5 minutes' standing, 10 minutes' low-intensity cycling at 50 Watts and 10 minutes' moderate-intensity cycling at 100 Watts. The trial was repeated on two different days, with a median of 110 days in between. Continuous 64 Hz finger PPG was collected by Shimmer GSR+, while beat-to-beat CO was estimated by the Finapres device using the Modelflow® pulse-contour analysis of the continuous finger ABPM. PPG was synchronized with ABPW via the system time stamps. Linear detrending, a 0.5-second median filter and a 4th-order Butterworth bandpass filter were applied to correct baseline drift, suppress motion artifact and reduce noise.

To apply this dataset for model evaluation, each recording was normalized into (0,1). Similarly, the median value of the corresponding CO values of each PPG segment was chosen as the label. From the measurement, the upper limit for CO values is 16.4 L/min, which conforms to gated blood pool scans based CO observations in subjects at rest and during 100 Watts' bicycle exercise \cite{Rodeheffer1984}. Then, the segment with CO values less than 4.3 L/min are considered implausible in exercise physiology according to the lower limit in 20-30 aged group \cite{Rusinaru2021}. 

\subsubsection{Dataset window construction}

Each dataset was segmented into 5s, 15s, and 30s PPG segments using half-duration sliding windows to evaluate the model’s performance across various practical scenarios. The 5-second segments provide sufficient pulse waveform features for capturing heart rate and respiration-related variations, while maintaining low computational cost, which is ideal for real-time applications in wearable and mobile health devices \cite{Parak2014, Icenhower2025}. The 15-second segments offer a balanced trade-off between information richness and computational efficiency, leveraging multiple cardiac cycles for stable cardiovascular dynamics estimation in a practice common to both clinical and commercial monitoring \cite{Wang2023, Lingawi2025}. The 30-second segments are capable of capturing longer physiological trends and more complex hemodynamic patterns, a critical requirement for attention-based generative models validated in recent studies like GPT-PPG \cite{Chen2025GPTPPG}. Simultaneously, by encompassing a sufficient number of cardiac cycles, these segments enhance the statistical robustness of feature extraction, thereby effectively mitigating measurement deviations caused by a single, low-quality PPG beat. A summary of the three datasets is provided in Supplementary Table S1, where the data amounts refer to the number of subjects and segments remaining after removal of physiologically implausible segment-level CO labels.

\subsection{Experimental setting}
CVAF-Net was implemented using Pytorch 2.1.0. All experiments were conducted on an NVIDIA GeForce RTX 2080ti GPU. To ensure the reproducibility of the proposed CVAF-Net, the detailed hyperparameters for each component and the training configuration are summarized in Supplementary Table S2.

Training used the Adam optimizer (learning rate$ = $1e-4, weight decay$ = $1e-5), MSE loss, batch size$ = $32 and a fixed number of epochs$ = $100. Model complexity (FLOPs and parameter count) was estimated with THOP. During training the best validation model per fold was checkpointed and later loaded for evaluation.

To comprehensively evaluate the strengths and limitations of the proposed CVAF-Net against comparative baseline methods, we implemented a subject-wise 5-fold cross-validation strategy to ensure robustness and minimize sampling bias. Specifically, the data partitioning was performed at the subject level to strictly prevent data leakage. For each fold, subjects were split into training, validation, and testing sets according to a 7:1:2 ratio, ensuring that segments from the same subject never appeared in different sets simultaneously. Final performance metrics were determined by averaging the results across all five folds.

\subsection{Baseline methods}
To comprehensively evaluate the performance of CVAF-Net, multiple methods including traditional regression and state-of-the-art deep learning methods were included in comparison. 
\begin{enumerate}
    \item Traditional regression methods: support vector regression (SVR), random forest regression (RFR) and extreme gradient boosting (XGBoost) treated the PPG temporal signal as input. 
    \item Deep learning methods: fully-connected neural network (FCNN), long short-term memory (LSTM) treated temporal signal as input. CNN-LSTM and ResNet-18 utilized the 8-channel FSM as input, while ResNet-18 was additionally evaluated using a 9-channel map formed by concatenating the PPG temporal signal with the FSM. Transformer processes the temporal signal and feature sequence map through separate 1D convolutional layers, which are then concatenated along the feature dimension to form a unified sequence for global cross-modal modeling.
\end{enumerate}

\subsection{Model evaluation}
\subsubsection{Performance evaluation}
\label{3.4.1}
To evaluate the estimation accuracy of CVAF-Net and all the other baseline methods, four commonly used evaluation metrics, R squared score, mean absolute error (MAE), root mean square error (RMSE) and mean absolute percentage error (MAPE) were applied. The calculation formulas of these metrics are:
\begin{equation}
R^2 = 1 - \frac{\sum_{i=1}^{n}(y_i - \hat{y}_i)^2}
{\sum_{i=1}^{n}(y_i - \bar{y})^2}
\end{equation}

\begin{equation}
\mathrm{MAE} = \frac{1}{n} \sum_{i=1}^{n} |y_i - \hat{y}_i|
\end{equation}

\begin{equation}
\mathrm{RMSE} =
\sqrt{
\frac{1}{n}
\sum_{i=1}^{n}(y_i - \hat{y}_i)^2
}
\end{equation}

\begin{equation}
\mathrm{MAPE} =
\frac{1}{n}
\sum_{i=1}^{n}
\left|
\frac{y_i - \hat{y}_i}{y_i}
\right|
\times 100
\end{equation}
where $y_i$ and $\hat{y}_i$ denotes the true and estimated value of cardiac output and n denotes the sample amount of the testing set. Furthermore, a Wilcoxon signed-rank test was performed to evaluate the statistical significance of the subject-wise MAE differences between each baseline model and CVAF-Net. A $p$-value $< 0.05$ was considered statistically significant. Computational efficiency and model complexity of CVAF-Net were evaluated against baseline architectures in terms of floating-point operations (FLOPs) and parameter counts across the three window scenarios.

\subsubsection{Ablation study}
To verify the contribution of each module in CVAF-Net, we conducted an ablation study by systematically removing or modifying individual modules. Specifically, we designed the following three variants:
\begin{enumerate}
    \item I: No spatial encoder. In this variant, only the temporal encoder was retained to capture temporal dynamics.
    \item II: No temporal encoder. In this variant, only the spatial encoder was retained to capture FSM-based spatial dynamics.
    \item III: No cross view attention fusion. In this variant, the outputs of the spatial and temporal encoders were individually processed by average pooling, then concatenated and passed through a fully connected layer for regression.
\end{enumerate}

Performance of each variant was evaluated with the same four metrics in \ref{3.4.1}. Wilcoxon signed-rank test was performed to evaluate the statistical significance of the subject-wise MAE differences between each variant and CVAF-Net. The sample size for the comparison was the same in \ref{3.4.1} for each dataset. A $p$-value $< 0.05$ was considered statistically significant.

\subsubsection{Plausibility analysis}
We analyzed the plausibility for predictions from all five cross-validation test folds, covering the entire subject pool, for each sequence length (5-second, 15-second, and 30-second) across the SPW and RC datasets. To ensure the clinical utility of the proposed CVAF-Net, it is essential to verify that the estimated CO aligns with established physiological principles. We conducted a correlation analysis across two distinct datasets to validate the plausibility of our model's predictions. Notably, the VP dataset was excluded from this analysis. During these specific physiological challenges, changes in CO are highly transient and governed by multiple interacting autonomic mechanisms. As a result, changes in CO cannot be reliably characterized by simple correlations with individual cardiovascular parameters, particularly due to rapid compensatory responses and abrupt alterations in venous return.
\begin{enumerate}
    \item SPW dataset: Using this simulated dataset containing virtual subjects, we examined the correlation between predicted CO and age. This allows us to verify if the model correctly captures the age-related decline in cardiovascular capacity across a broad demographic spectrum. From a physiological perspective, resting CO is expected to exhibit a significant inverse correlation with age, as aging is typically accompanied by a progressive decline in maximal heart rate and stroke volume. We calculated the median predicted CO for each age group and derived boxplots to observe the age-related trend. The Spearman's rank correlation coefficient ($\rho$) was computed across all individual data points to quantify the monotonic association between predicted CO and age.
    \item RC dataset: Using data from 10 real subjects during exercise, we analyzed the relationship between predicted CO and HR, as well as SVR. We expect to observe a strong positive correlation with HR and a strong negative correlation with SVR, validating the model's ability to track exercise-induced hemodynamic transitions. During physical exertion or exercise, CO must increase substantially to meet metabolic demands. This process is primarily driven by a concomitant rise in heart rate (HR). Furthermore, according to the hemodynamic relationship CO = MAP/SVR, we expect estimated CO to show a strong negative correlation with systemic vascular resistance (SVR), particularly as vasodilation during activity reduces SVR to facilitate higher blood flow. We examined the relationships between predicted CO and HR/SVR. Scatter plots were generated to visualize these correlations at the individual level. We calculated the Spearman's rank correlation coefficient ($\rho$) for each subject and reported the median $\rho$ across the 10 subjects for each scenario to assess the consistency of the model's physiological alignment.
\end{enumerate}

\subsubsection{Motion-artifact robustness analysis}
Robustness was assessed by perturbing each test PPG segment while retaining the original CO label and trained model. Four conditions were evaluated: clean PPG; level 1, comprising additive white Gaussian noise (AWGN) at 25~dB signal-to-noise ratio (SNR) and sinusoidal baseline wander with an amplitude of 0.05; level 2, comprising 15~dB AWGN and a baseline-wander amplitude of 0.15; and level 3, comprising 5~dB AWGN, a baseline-wander amplitude of 0.40, and transient burst noise. Baseline wander used a frequency of 0.2~Hz. Burst noise was generated from high-amplitude Gaussian spikes filtered by a fourth-order low-pass Butterworth filter with a 5~Hz cut-off.

Noise variance was determined from
\begin{equation}
\mathrm{SNR} = 10 \log_{10}\left(\frac{P_{\mathrm{signal}}}{P_{\mathrm{noise}}}\right).
\end{equation}
Perturbations were applied only to the PPG input. All derivative, Fourier, Hilbert, and wavelet channels were then regenerated from the perturbed waveform, ensuring that each model received a noise-consistent version of its required input. RMSE was calculated against the unchanged reference CO for every model, dataset, window length, and artifact level.

\section{Results}
\subsection{Performance and comparison}
Across the longitudinal analysis of different segment lengths, CVAF-net exhibited exceptional stability and robustness, as evidenced by the minimal performance degradation when reducing window size, which underscores the effectiveness of the dual-branch architecture and cross-view attention fusion in capturing robust hemodynamic features without reliance on explicit pulse cycle detection. The comparative evaluation across three distinct datasets, shown in Table \ref{tab:performance_results}, where bold and italics represent the first- and second-best performance, demonstrates the consistent superiority of the proposed CVAF-net against various state-of-the-art baselines for CO estimation from various PPG segments. On the simulated SPW dataset, CVAF-Net achieved the best performance in the 15-second setting (MAE = 0.19 L/min, MAPE = 3.95\%) and competitive performance in the 30-second setting (MAE = 0.25 L/min, MAPE = 5.36\%), while the Transformer performed best in the 5-second setting (MAE = 0.20 L/min, MAPE = 4.26\%). On the more challenging RC and VP datasets, CVAF-Net generally ranked among the best-performing models and achieved the lowest MAE in several settings; however, its margin over the Transformer was modest in some comparisons. For example, on the RC dataset, CVAF-Net achieved its lowest MAE of 1.58 L/min at 15 seconds, compared with 1.64 L/min for the Transformer. On the VP dataset, CVAF-Net achieved a minimum MAE of 1.20 L/min at 5 seconds, compared with 1.27 L/min for the Transformer.

\begin{table*}[htbp]
\centering
\footnotesize 
\setlength{\tabcolsep}{5.5pt}
\caption{Performance comparison across datasets and time windows. Bold and italic values indicate the best and second-best results, respectively. $^{\dagger}$ indicates a significant difference ($p < 0.05$) between the corresponding model and CVAF-Net.}
\label{tab:performance_results}
\begin{tabular}{llcccccccccccc}
\toprule
\multirow{2}{*}{Model} & \multirow{2}{*}{Input} & \multicolumn{4}{c}{5-second} & \multicolumn{4}{c}{15-second} & \multicolumn{4}{c}{30-second} \\ \cmidrule(lr){3-6} \cmidrule(lr){7-10} \cmidrule(lr){11-14}
&  & $R^2$ & MAE & RMSE & MAPE & $R^2$ & MAE & RMSE & MAPE & $R^2$ & MAE & RMSE & MAPE \\ \midrule
\multicolumn{14}{l}{\textbf{SPW dataset}} \\
SVR & TS & 0.45 & 0.61$^{\dagger}$ & 0.80 & 13.85 & 0.56 & 0.56$^{\dagger}$ & 0.72 & 12.53 & 0.35 & 0.52$^{\dagger}$ & 0.66 & 10.94 \\
RFR & TS & 0.55 & 0.59$^{\dagger}$ & 0.73 & 13.61 & 0.62 & 0.55$^{\dagger}$ & 0.67 & 12.67 & 0.44 & 0.50$^{\dagger}$ & 0.61 & 10.78 \\
XGBoost & TS & 0.68 & 0.46$^{\dagger}$ & 0.62 & 10.67 & 0.76 & 0.40$^{\dagger}$ & 0.53 & 9.22 & 0.60 & 0.39$^{\dagger}$ & 0.51 & 8.35 \\
FCNN & TS & 0.88 & 0.30$^{\dagger}$ & 0.37 & 6.81 & 0.86 & 0.32$^{\dagger}$ & 0.40 & 7.33 & 0.78 & 0.30$^{\dagger}$ & 0.37 & 6.19 \\
LSTM & TS & 0.50 & 0.64$^{\dagger}$ & 0.77 & 15.29 & 0.43 & 0.69$^{\dagger}$ & 0.83 & 16.17 & 0.52 & 0.63$^{\dagger}$ & 0.75 & 14.70 \\
CNN-LSTM & SFM & 0.83 & 0.33$^{\dagger}$ & 0.45 & 7.63 & 0.88 & 0.24$^{\dagger}$ & 0.37 & 5.74 & 0.65 & 0.33$^{\dagger}$ & 0.45 & 7.11 \\
ResNet-18 & SFM & 0.89 & 0.25$^{\dagger}$ & 0.36 & 5.69 & 0.90 & 0.23$^{\dagger}$ & 0.34 & 5.21 & 0.65 & 0.35$^{\dagger}$ & 0.45 & 7.44 \\
ResNet-18 & TS + SFM & 0.89 & 0.26$^{\dagger}$ & 0.37 & 5.74 & \textit{0.91} & \textit{0.23}$^{\dagger}$ & \textit{0.32} & \textit{5.12} & 0.67 & 0.33$^{\dagger}$ & 0.43 & 6.97 \\
Transformer & TS + SFM & \textbf{0.94} & \textbf{0.20}$^{\dagger}$ & \textbf{0.27} & \textbf{4.67} & 0.84 & 0.33$^{\dagger}$ & 0.41 & 7.75 & \textit{0.83} & \textit{0.28}$^{\dagger}$ & \textit{0.39} & \textit{6.54} \\
CVAF-Net & TS + SFM & \textit{0.92} & \textit{0.23} & \textit{0.30} & \textit{5.15} & \textbf{0.93} & \textbf{0.19} & \textbf{0.27} & \textbf{3.95} & \textbf{0.87} & \textbf{0.25} & \textbf{0.34} & \textbf{5.50} \\ \midrule
\multicolumn{14}{l}{\textbf{RC dataset}} \\
SVR & TS & 0.23 & 1.81$^{\dagger}$ & 2.30 & 22.31 & 0.20 & 1.88$^{\dagger}$ & 2.38 & 23.52 & 0.16 & 1.90$^{\dagger}$ & 2.40 & 24.08 \\
RFR & TS & 0.22 & 1.86$^{\dagger}$ & 2.31 & 23.99 & 0.25 & 1.84$^{\dagger}$ & 2.30 & 23.89 & 0.21 & 1.88$^{\dagger}$ & 2.33 & 24.39 \\
XGBoost & TS & 0.24 & 1.84$^{\dagger}$ & 2.33 & 23.78 & 0.26 & 1.81$^{\dagger}$ & 2.29 & 23.54 & 0.22 & 1.86$^{\dagger}$ & 2.32 & 23.94 \\
FCNN & TS & 0.11 & 1.98$^{\dagger}$ & 2.50 & 25.23 & 0.12 & 2.00$^{\dagger}$ & 2.47 & 26.64 & 0.07 & 2.05$^{\dagger}$ & 2.53 & 27.84 \\
LSTM & TS & 0.17 & 1.94$^{\dagger}$ & 2.53 & 24.63 & 0.15 & 1.89$^{\dagger}$ & 2.41 & 25.36 & 0.12 & 1.90$^{\dagger}$ & 2.49 & 27.18 \\
CNN-LSTM & SFM & 0.20 & 1.86$^{\dagger}$ & 2.37 & 23.76 & 0.16 & 1.80$^{\dagger}$ & 2.34 & 23.36 & 0.27 & 1.84$^{\dagger}$ & 2.31 & 24.92 \\
ResNet-18 & SFM & 0.21 & 1.80$^{\dagger}$ & 2.33 & 22.96 & 0.22 & 1.74$^{\dagger}$ & 2.24 & 22.79 & 0.25 & 1.71$^{\dagger}$ & 2.18 & 22.46 \\
ResNet-18 & TS + SFM & 0.21 & 1.79$^{\dagger}$ & 2.36 & 22.72 & 0.24 & 1.74$^{\dagger}$ & 2.24 & 22.33 & 0.29 & 1.69$^{\dagger}$ & 2.16 & 21.40 \\
Transformer & TS + SFM & \textit{0.30} & \textit{1.67} & \textit{2.15} & \textit{20.63} & \textit{0.28} & \textit{1.64} & \textit{2.13} & \textit{20.25} & \textit{0.33} & \textit{1.64} & \textit{2.08} & \textit{20.49} \\
CVAF-Net & TS + SFM & \textbf{0.38} & \textbf{1.60} & \textbf{1.98} & \textbf{18.70} & \textbf{0.36} & \textbf{1.58} & \textbf{1.99} & \textbf{19.15} & \textbf{0.39} & \textbf{1.59} & \textbf{2.01} & \textbf{19.49} \\ \midrule
\multicolumn{14}{l}{\textbf{VP dataset}} \\
SVR & TS & -0.05 & 1.52$^{\dagger}$ & 1.98 & 30.88 & -0.03 & 1.49$^{\dagger}$ & 1.96 & 30.02 & -0.07 & 1.72$^{\dagger}$ & 2.24 & 32.54 \\
RFR & TS & 0.05 & 1.44$^{\dagger}$ & 1.88 & 29.10 & 0.07 & 1.42$^{\dagger}$ & 1.86 & 28.48 & 0.03 & 1.63$^{\dagger}$ & 2.12 & 33.54 \\
XGBoost & TS & 0.15 & 1.36$^{\dagger}$ & 1.78 & 27.66 & 0.08 & 1.45$^{\dagger}$ & 1.75 & 26.39 & 0.12 & 1.55$^{\dagger}$ & 2.01 & 29.62 \\
FCNN & TS & -0.02 & 1.48$^{\dagger}$ & 1.95 & 29.73 & 0.00 & 1.50$^{\dagger}$ & 1.92 & 30.61 & -0.04 & 1.53$^{\dagger}$ & 2.06 & 30.11 \\
LSTM & TS & 0.07 & 1.50$^{\dagger}$ & 1.93 & 30.25 & 0.10 & 1.46$^{\dagger}$ & 1.87 & 29.88 & 0.09 & 1.59$^{\dagger}$ & 2.17 & 32.16 \\
CNN-LSTM & SFM & 0.06 & 1.34$^{\dagger}$ & 1.85 & 27.25 & -0.01 & 1.45$^{\dagger}$ & 1.93 & 31.11 & -0.04 & 1.60$^{\dagger}$ & 2.08 & 33.71 \\
ResNet-18 & SFM & 0.12 & 1.35$^{\dagger}$ & 1.79 & 27.62 & -0.07 & 1.42$^{\dagger}$ & 1.89 & 29.64 & 0.00 & 1.57$^{\dagger}$ & 1.89 & 31.77 \\
ResNet-18 & TS + SFM & 0.13 & 1.36$^{\dagger}$ & 1.80 & 27.70 & -0.08 & 1.44$^{\dagger}$ & 1.91 & 29.71 & 0.01 & 1.56$^{\dagger}$ & 1.90 & 31.82 \\
Transformer & TS + SFM & \textit{0.19} & \textit{1.27} & \textit{1.74} & \textit{24.61} & \textit{0.08} & \textit{1.27} & \textit{1.68} & \textit{25.02} & \textit{0.10} & \textit{1.39} & \textit{1.85} & \textit{29.88} \\
CVAF-Net & TS + SFM & \textbf{0.25} & \textbf{1.20} & \textbf{1.60} & \textbf{22.83} & \textbf{0.18} & \textbf{1.24} & \textbf{1.48} & \textbf{20.62} & \textbf{0.19} & \textbf{1.34} & \textbf{1.71} & \textbf{26.31} \\ \bottomrule
\end{tabular}
\end{table*}

CVAF-Net achieves an effective balance between representational capacity and computational demand, making it a viable candidate for deployment in high-performance wearable health monitoring systems. Model complexity results are reported in Table \ref{tab:model_complexity} to complement the accuracy comparison. CVAF-Net has a parameter count of 5.115M since it includes both a ResNet-based spatial encoder and an LSTM-based temporal encoder. However, compared with the Transformer, which achieved similarly strong estimation accuracy, CVAF-Net required substantially fewer FLOPs. In the 15-second scenario, for example, CVAF-Net required 245.16M FLOPs, whereas the Transformer required 3.04G FLOPs, corresponding to a more than twelvefold reduction in computational operations. This comparison suggests that CVAF-Net provides a more favorable accuracy-computation trade-off than the Transformer baseline, even though simpler models such as FCNN and LSTM remain lighter but less accurate.
\begin{table*}[htbp]
\centering
\footnotesize 
\caption{Model Complexity Comparison across Different Time Scenarios}
\label{tab:model_complexity}
\begin{tabular}{lcccccc}
\toprule
\multirow{2}{*}{Model} & \multicolumn{2}{c}{5-second} & \multicolumn{2}{c}{15-second} & \multicolumn{2}{c}{30-second} \\ 
\cmidrule(lr){2-3} \cmidrule(lr){4-5} \cmidrule(lr){6-7}
& FLOPs & Parameter & FLOPs & Parameter & FLOPs & Parameter \\ 
\midrule
FCNN & 251.20K & 125.441K & 581.44K & 125.441K & 1.08M & 125.441K \\
LSTM & 84.56M & 209.537K & 253.64M & 209.537K & 507.27M & 209.537K \\
CNN-LSTM & 54.76M & 1.877M & 164.21M & 1.877M & 328.38M & 1.877M \\
Resnet-18 & 55.57M & 4.011M & 166.38M & 4.011M & 332.59M & 4.011M \\
Transformer & 1.01G & 3.269M & 3.04G & 3.269M & 6.09G & 3.269M \\
CVAF-Net & 76.782M & 5.115M & 245.16M & 5.115M & 472.36M & 5.115M \\
\bottomrule
\end{tabular}
\end{table*}

\subsection{Ablation study}
The ablation study results, summarized in Table \ref{tab:ablation}, systematically validate the individual contributions of each architectural component within the CVAF-Net framework. The removal of the spatial encoder (Variant I) leads to the most significant performance degradation across all datasets, particularly in the VP scenario where R$^2$ values drop to near-zero levels, highlighting the critical necessity of the multi-channel feature sequence map for robust CO estimation. While the spatial-only configuration (Variant II) outperforms the temporal-only variant, it remains consistently inferior to the full CVAF-Net, confirming that raw 1D temporal dynamics provide essential complementary information that cannot be captured by transformed features alone. Furthermore, the performance gap between Variant III (simple concatenation) and the full model demonstrates the superiority of the cross-view attention fusion mechanism over traditional fusion methods. By intelligently modeling the complex interdependencies between spatial and temporal representations, the complete CVAF-Net achieves the lowest error metrics across all time windows and datasets.
\begin{table*}[htbp]
\centering
\footnotesize
\caption{Performance Evaluation for Different Ablation Scenarios. $^{\dagger}$ reveals significant difference ($p < 0.05$) between the current model and the CVAF-Net.}
\label{tab:ablation}
\begin{tabular}{clcccccccccccc}
\toprule
\multirow{2}{*}{Dataset} & \multirow{2}{*}{Variant} & \multicolumn{4}{c}{5-second} & \multicolumn{4}{c}{15-second} & \multicolumn{4}{c}{30-second} \\ \cmidrule(lr){3-6} \cmidrule(lr){7-10} \cmidrule(lr){11-14}
&  & $R^2$ & MAE & RMSE & MAPE & $R^2$ & MAE & RMSE & MAPE & $R^2$ & MAE & RMSE & MAPE \\ \midrule
\multirow{4}{*}{SPW} 
 & I & 0.47 & 0.66$^{\dagger}$ & 0.84 & 16.38 & 0.44 & 0.62$^{\dagger}$ & 0.81 & 16.22 & 0.43 & 0.66$^{\dagger}$ & 0.89 & 15.98 \\
 & II & 0.86 & 0.28$^{\dagger}$ & 0.39 & 5.97 & 0.87 & 0.27$^{\dagger}$ & 0.38 & 6.01 & 0.67 & 0.36$^{\dagger}$ & 0.51 & 7.55 \\
 & III & 0.88 & 0.30$^{\dagger}$ & 0.38 & 6.85 & 0.88 & 0.26$^{\dagger}$ & 0.37 & 5.96 & 0.71 & 0.36$^{\dagger}$ & 0.47 & 7.29 \\
 & CVAF-Net & 0.92 & 0.23 & 0.30 & 5.15 & 0.93 & 0.19 & 0.27 & 3.95 & 0.87 & 0.25 & 0.34 & 5.50 \\ \midrule
 \multirow{4}{*}{RC} 
 & I & 0.14 & 1.90$^{\dagger}$ & 2.49 & 25.12 & 0.12 & 1.93$^{\dagger}$ & 2.83 & 26.04 & 0.14 & 1.86$^{\dagger}$ & 2.53 & 25.87 \\
 & II & 0.26 & 1.73$^{\dagger}$ & 2.30 & 21.73 & 0.22 & 1.78$^{\dagger}$ & 2.27 & 22.58 & 0.24 & 1.75$^{\dagger}$ & 2.24 & 21.89 \\
 & III & 0.25 & 1.76$^{\dagger}$ & 2.23 & 22.62 & 0.34 & 1.60$^{\dagger}$ & 2.08 & 20.41 & 0.29 & 1.68$^{\dagger}$ & 2.19 & 21.96 \\
 & CVAF-Net & 0.38 & 1.60 & 1.98 & 18.70 & 0.36 & 1.58 & 1.99 & 19.15 & 0.39 & 1.59 & 2.01 & 19.49 \\ \midrule
 \multirow{4}{*}{VP}  
 & I & 0.07 & 1.52$^{\dagger}$ & 1.89 & 30.63 & 0.04 & 1.61$^{\dagger}$ & 1.97 & 30.10 & 0.01 & 1.71$^{\dagger}$ & 2.33 & 33.62 \\
 & II & 0.20 & 1.33$^{\dagger}$ & 1.83 & 24.89 & 0.15 & 1.43$^{\dagger}$ & 1.88 & 26.01 & 0.06 & 1.59$^{\dagger}$ & 2.02 & 31.24 \\
 & III & 0.21 & 1.30$^{\dagger}$ & 1.72 & 24.67 & 0.10 & 1.37$^{\dagger}$ & 1.83 & 25.86 & 0.09 & 1.49$^{\dagger}$ & 1.94 & 30.02 \\
 & CVAF-Net & 0.25 & 1.20 & 1.60 & 22.83 & 0.18 & 1.24 & 1.48 & 20.62 & 0.19 & 1.34 & 1.71 & 26.31 \\ \bottomrule
\end{tabular}
\end{table*}

\subsection{Plausibility analysis}

Across the physiological plausibility analyses, CVAF-Net consistently ranked among the top-performing models across window lengths, indicating that its strong predictive accuracy was accompanied by preservation of physiologically meaningful hemodynamic relationships. This consistency was observed for both the population-level association between CO and age in SPW and the subject-level associations of CO with HR and SVR in RC (Supplementary Figure S1).

On SPW, every model produced an inverse association between predicted CO and age. Across the baseline models, Spearman's ($\rho$) ranged from -0.399 to -0.284 at 5 s, from -0.372 to -0.253 at 15 s, and from -0.278 to -0.134 at 30 s. CVAF-Net yielded ($\rho$) values of -0.367, -0.362, and -0.274, respectively. Thus, although the strongest coefficient varied between models at individual window lengths, CVAF-Net remained within the leading group under all three conditions and showed particularly stable preservation of the expected inverse age-CO relationship as the window length changed.

A similar pattern was observed for the activity-related hemodynamic relationships in RC. CVAF-Net produced positive median CO-HR correlations of 0.784, 0.798, and 0.894 at 5 s, 15 s, and 30 s, respectively, with the 30 s correlation being the strongest among the compared models. Its median CO-SVR correlations were consistently inverse, at -0.724, -0.740, and -0.739 across the three window lengths. Although several baseline models also preserved these expected relationships, CVAF-Net remained among the strongest models for both associations and across all window lengths.

\subsection{Motion-artifact robustness}

CVAF-Net consistently remained within the top-performing group under progressively stronger motion-artifact perturbations. Its competitive performance was therefore maintained under both clean and degraded PPG conditions. Across datasets and window lengths, CVAF-Net and the Transformer generally formed the lowest-error group, whereas several alternative models showed substantially greater degradation (Supplementary Figure S2).

For CVAF-Net, the ranges across the 5-, 15-, and 30-s windows increased progressively at each artifact level (all values in L/min). In SPW, RMSE ranged from 0.27–0.34 under clean conditions, 0.30–0.38 at level 1, 0.36–0.48 at level 2, and 0.51–0.65 at level 3. In RC, the corresponding ranges were 1.98–2.01, 2.16–2.21, 2.41–2.66, and 3.18–3.66. In VP, RMSE increased from 1.48–1.71 under clean conditions to 1.74–1.87 at level 1, 2.02–2.16 at level 2, and 2.69–3.11 at level 3. The increase was relatively modest from clean PPG to level 1, became more pronounced at level 2, and was largest at level 3. Despite the overall degradation, model ranking remained broadly stable across the nine dataset-window combinations. CVAF-Net consistently remained among the models with the lowest RMSE, together with the Transformer.

\section{Discussion}
This study introduces CVAF-Net, a prior-guided, dual-view deep learning architecture designed to estimate cardiac output (CO) from short-duration photoplethysmography (PPG) segments without relying on explicit pulse-cycle segmentation. Our central finding demonstrates that decoupling the raw temporal waveform from a transformed feature sequence map (FSM), and subsequently orchestrating their interaction via cross-view attention, yields highly competitive CO estimation across simulated, exercise-induced, and vasoconstrictive hemodynamic states. Crucially, CVAF-Net achieves this predictive accuracy while offering a twelvefold reduction in computational complexity compared with the Transformer baseline. Collectively, these findings support our central hypothesis that CO-related information is more effectively captured when raw temporal dynamics and prior-guided signal representations are learned as complementary views and selectively integrated through cross-view attention.

\subsection{Architectural contribution and predictive performance}
The mapping from peripheral optical pulsations to central CO is indirect and is influenced by stroke volume, vascular tone, arterial compliance, and local perfusion \cite{Charlton2022}. Raw PPG preserves continuous timing and morphology, while the FSM emphasizes slope, spectral, envelope, and multiscale structure. The ablation ordering supports their complementarity. The temporal-only model lost the structured view, whereas the spatial-only model lost untransformed dynamics. Their inferior performance indicates that the two views provide complementary information for CO estimation. Pooled concatenation retained both representations but lacked the conditional interaction implemented by cross-view attention. Its lower performance suggests that the gain of the complete architecture depends not only on the availability of complementary representations, but also on learning how information from one view should be weighted in the context of the other. These findings answer the central hypothesis at the performance level. Identifying the specific latent features responsible for the gain requires attribution and representation analyses.

The performance gap between simulated and measured data was substantial. SPW provided a controlled mapping with noiseless waveforms and exact simulated labels, enabling RMSE values below 0.4 L/min for the strongest models. RC and VP include sensor noise, motion, contact-pressure variation, autonomic responses, and pulse-contour reference error. These factors introduce variability in both the PPG input and the reference labels, weakening the observable relation between peripheral PPG morphology and central CO. Accordingly, SPW tests the architecture under controlled conditions, whereas RC and VP provide a more demanding assessment under measurement and physiological variability. The higher errors on RC and VP therefore define the present generalization boundary more clearly than the SPW results alone.

\subsection{Physiological plausibility across model families}
The SPW analysis showed weak-to-moderate inverse associations between predicted CO and age for every model. CVAF-Net was among the stronger models at all window lengths, although XGBoost and the Transformer reached similar or stronger coefficients in individual settings. These coefficients represent marginal population-level associations because age does not uniquely determine CO. In the simulated cohort, concurrent variation in heart rate, stroke volume, vascular tone, and arterial compliance may attenuate the monotonic association with age \cite{Lakatta2003,Strait2012}. Moreover, Spearman’s $\rho$ is unadjusted and cannot isolate age from these interacting determinants. The inverse correlation therefore provides a complementary check of physiological consistency, but does not explain prediction accuracy or indicate that age drives individual estimates.

In RC, CVAF-Net retained the expected exercise-related directions: CO rose with HR and decreased with SVR. Its CO-HR association strengthened at 30 s, reaching a median rho of 0.894, while the CO-SVR association remained stable near -0.74. Several baselines showed comparable SVR associations, indicating that physiological concordance was not exclusive to CVAF-Net. Nevertheless, plausibility is more informative when considered jointly with estimation accuracy. A model may reproduce the direction of a broad physiological trend while still yielding inaccurate absolute estimates, whereas low aggregate error alone does not guarantee coherent behavior across changing physiological states. CVAF-Net preserved these associations while maintaining the lowest or near-lowest prediction error, strengthening the evidence that its predictions remain consistent with the underlying hemodynamic changes represented in the data.

\subsection{Sensitivity to motion artifacts}
The graded stress test exposed a limitation hidden by clean-signal metrics. RMSE increased with each artifact level, and the largest increment appeared when 5 dB AWGN, strong baseline wander, and burst noise were combined. This pattern is plausible because severe transient artifacts distort local morphology and propagate into every FSM channel regenerated from the corrupted waveform. Longer windows provided more context but did not eliminate sensitivity, indicating that additional temporal information cannot fully compensate for severe corruption. Robustness therefore depends on signal quality as well as sequence length, which is particularly relevant for ambulatory PPG where motion and sensor-contact changes can occur intermittently.

CVAF-Net and the Transformer formed the most robust group, although their errors still increased substantially. The LSTM was generally most sensitive, consistent with a single temporal representation being disrupted by local amplitude and baseline changes. Feature-based regressors also degraded, showing that explicit signal processing does not guarantee artifact tolerance. The competitive robustness of CVAF-Net suggests that its dual-view representation remains informative under moderate perturbation, although the present experiment cannot determine which branch or fusion operation contributes most to this behavior. These results motivate signal-quality gating, artifact-aware augmentation, and uncertainty estimation before free-living use.

\subsection{Computational and translational implications}
CVAF-Net used more parameters than the simplest baselines but required far fewer operations than the Transformer. The greater than twelvefold FLOP reduction at 15 s is relevant to edge inference, where memory, power, and latency are constrained. Repeated short-window inference can amplify computational demands in continuous monitoring. Fixed-duration inputs also support rolling estimates without explicit beat segmentation, avoiding a preprocessing dependency that may become unreliable when individual pulses are poorly delineated. Together, these properties could enable frequent hemodynamic trend updates on wearable or near-sensor hardware while retaining richer representations than lightweight single-branch models.

However, FLOPs and parameter counts are architecture-level indicators instead of direct measurements of deployment performance. Actual latency, energy consumption, and memory use will depend on the target hardware and implementation. The current evidence therefore supports trend monitoring and algorithm development, not replacement of invasive CO measurement. A short-window model could help identify sustained deviations or delayed recovery in monitored patients, provided that signal quality and uncertainty are reported. Demonstrating timely detection of acute deterioration will require prospective studies with prespecified clinical endpoints and direct reference measurements.

\subsection{Limitations and future work}
Several limitations constrain interpretation. First, SPW segments were created by repeating simulated beats, so they do not reproduce natural beat-to-beat variability. Consequently, longer SPW windows provide repeated waveform context instead of the full temporal variability expected in continuous recordings. Second, the measured cohorts were small and demographically restricted, especially RC with 10 participants. Third, RC and VP reference CO values were derived from pulse-contour analysis and include method-specific uncertainty. Fourth, synthetic artifacts approximate common PPG disturbances but cannot reproduce the full coupling between body motion, sensor contact, tissue optics, and physiology. Fifth, models were evaluated within each dataset; transfer to unseen devices, acquisition protocols, sites, and patient groups remains unknown. These limitations are particularly relevant because wearable deployment may introduce several of these distribution shifts simultaneously.

Plausibility correlations are post-hoc associations and do not identify the information used by the network. The current results therefore support physiological consistency of the outputs but not mechanistic interpretability of the learned representation. Future work should combine external multicenter validation with direct or clinically accepted reference measurements, device-shift testing, prospective motion protocols, calibrated uncertainty, and channel-level attribution. Cross-dataset evaluation would further help determine whether the dual-view representation captures transferable PPG-CO relationships. These studies are needed to determine whether the observed accuracy-computation balance persists in free-living cardiovascular populations.

\section{Conclusion}
CVAF-Net integrates raw temporal PPG with a structured multiscale signal view through cross-view attention. Across three datasets and three window lengths, the complete model achieved consistently low CO estimation error, outperformed its ablated variants, retained physiologically concordant associations, and remained among the strongest models under graded motion artifacts. Its computational demand was markedly lower than the Transformer baseline. These results support prior-guided dual-view learning for short-window PPG analysis. External validation, direct reference measurements, and prospective free-living testing remain essential before clinical use.

\bibliographystyle{IEEEtran}
\bibliography{main}

@misc{WHO2025,
   author = {{World Health Organization}},
   title = {Cardiovascular diseases (CVDs)},
   url = {https://www.who.int/news-room/fact-sheets/detail/cardiovascular-diseases-(cvds)},
   month = {7},
   year = {2025},
}

@article{Roth2020,
   author = {{G. A. Roth, G. A. Mensah, C. O. Johnson, et al.}},
   doi = {10.1016/j.jacc.2020.11.010},
   journal = {Journal of the American College of Cardiology},
   month = {12},
   number = {25},
   pages = {2982-3021},
   title = {Global Burden of Cardiovascular Diseases and Risk Factors, 1990-2019},
   volume = {76},
   year = {2020},
}

@article{Rusinaru2021,
   author = {{D. Rusinaru, Y. Bohbot, F. Djelaili, et al.}},
   doi = {10.1016/j.amjcard.2020.10.046},
   journal = {The American Journal of Cardiology},
   month = {2},
   pages = {128-133},
   title = {Normative Reference Values of Cardiac Output by Pulsed-Wave Doppler Echocardiography in Adults},
   volume = {140},
   year = {2021},
}

@article{Vahdatpour2019,
   author = {{C. Vahdatpour, D. Collins, S. Goldberg}},
   doi = {10.1161/jaha.119.011991},
   journal = {Journal of the American Heart Association},
   month = {4},
   number = {8},
   pages = {e011991},
   title = {Cardiogenic Shock},
   volume = {8},
   year = {2019},
}

@article{Geerts2011,
   author = {{B. F. Geerts, L. P. Aarts, J. R. Jansen}},
   doi = {10.1111/j.1365-2125.2010.03798.x},
   journal = {British Journal of Clinical Pharmacology},
   month = {3},
   number = {3},
   pages = {316-330},
   title = {Methods in pharmacology: measurement of cardiac output},
   volume = {71},
   year = {2011},
}

@article{Grensemann2018,
   author = {{J. Grensemann}},
   doi = {10.3389/fmed.2018.00064},
   journal = {Frontiers in Medicine},
   month = {3},
   pages = {64},
   title = {Cardiac output monitoring by pulse contour analysis, the technical basics of less-invasive techniques},
   volume = {5},
   year = {2018},
}

@article{Saugel2018,
   author = {{B. Saugel, J. L. Vincent}},
   doi = {10.1097/mcc.0000000000000492},
   journal = {Current Opinion in Critical Care},
   month = {6},
   number = {3},
   pages = {165-172},
   title = {Cardiac output monitoring: how to choose the optimal method for the individual patient},
   volume = {24},
   year = {2018},
}

@article{Charlton2022,
   author = {{P. H. Charlton, P. A. Kyriacou, J. Mant, V. Marozas, P. Chowienczyk, J. Alastruey}},
   doi = {10.1109/jproc.2022.3149785},
   journal = {Proceedings of the IEEE},
   month = {3},
   number = {3},
   pages = {355-381},
   title = {Wearable photoplethysmography for cardiovascular monitoring},
   volume = {110},
   year = {2022},
}

@article{Elgendi2012,
   author = {{M. Elgendi}},
   doi = {10.2174/157340312801215782},
   journal = {Current Cardiology Reviews},
   month = {6},
   number = {1},
   pages = {14-25},
   title = {On the analysis of fingertip photoplethysmogram signals},
   volume = {8},
   year = {2012},
}

@article{Ipar2025,
   author = {{E. Ipar, L. J. Cymberknop, R. L. Armentano}},
   doi = {10.1088/1361-6579/adc366},
   journal = {Physiological Measurement},
   month = {3},
   number = {3},
   pages = {035008},
   title = {Parallel convolutional neural networks for non-invasive cardiac hemodynamic estimation: integrating uncalibrated PPG signals with nonlinear feature analysis},
   volume = {46},
   year = {2025},
}

@misc{Mol2020,
   author = {{A. Mol, C. G. M. Meskers, S. P. Niehof, A. B. Maier, R. van Wezel}},
   doi = {10.34973/te70-x603},
   publisher = {Radboud University},
   title = {Pulse transit time as a proxy for vasoconstriction in younger and older adults},
   year = {2020},
}

@article{Wang2023,
   author = {{W. Wang, P. Mohseni, K. L. Kilgore, L. Najafizadeh}},
   doi = {10.3389/fdgth.2022.1090854},
   journal = {Frontiers in Digital Health},
   month = {2},
   pages = {1090854},
   title = {PulseDB: A large, cleaned dataset based on MIMIC-III and VitalDB for benchmarking cuff-less blood pressure estimation methods},
   volume = {4},
   year = {2023},
}

@article{Chen2025GPTPPG,
   author = {{Z. Chen, C. Ding, S. Kataria, et al.}},
   doi = {10.48550/arXiv.2503.08015},
   journal = {arXiv preprint},
   month = {3},
   title = {GPT-PPG: A GPT-based foundation model for photoplethysmography signals},
   year = {2025},
   eprint = {2503.08015},
}

@article{Parak2014,
   author = {{J. Parak, I. Korhonen}},
   doi = {10.1109/embc.2014.6944419},
   journal = {2014 36th Annual International Conference of the IEEE Engineering in Medicine and Biology Society},
   month = {8},
   pages = {3670-3673},
   title = {Evaluation of wearable consumer heart rate monitors based on photopletysmography},
   year = {2014},
}

@article{Icenhower2025,
   author = {{A. Icenhower, C. Murphy, A. K. Brooks, et al.}},
   doi = {10.3389/fdgth.2025.1553565},
   journal = {Frontiers in Digital Health},
   month = {3},
   pages = {1553565},
   title = {Investigating the accuracy of Garmin PPG sensors on differing skin types based on the Fitzpatrick scale: cross-sectional comparison study},
   volume = {7},
   year = {2025},
}

@article{Tanaka1996,
   author = {{H. Tanaka, B. J. Sjöberg, O. Thulesius}},
   doi = {10.1111/j.1475-097x.1996.tb00565.x},
   journal = {Clinical Physiology},
   month = {3},
   number = {2},
   pages = {157-170},
   title = {Cardiac output and blood pressure during active and passive standing},
   volume = {16},
   year = {1996},
}

@article{Rodeheffer1984,
   author = {{R. J. Rodeheffer, G. Gerstenblith, L. C. Becker, J. L. Fleg, M. L. Weisfeldt, E. G. Lakatta}},
   doi = {10.1161/01.cir.69.2.203},
   journal = {Circulation},
   month = {2},
   number = {2},
   pages = {203-213},
   title = {Exercise cardiac output is maintained with advancing age in healthy human subjects: cardiac dilatation and increased stroke volume compensate for a diminished heart rate.},
   volume = {69},
   year = {1984},
}

@article{Lee2013,
   author = {{Q. Y. Lee, S. J. Redmond, G. S. Chan, et al.}},
   doi = {10.1186/1475-925x-12-19},
   journal = {BioMedical Engineering OnLine},
   month = {3},
   number = {1},
   pages = {19},
   title = {Estimation of cardiac output and systemic vascular resistance using a multivariate regression model with features selected from the finger photoplethysmogram and routine cardiovascular measurements},
   volume = {12},
   year = {2013},
}

@article{Wang2009,
   author = {{L. Wang, E. Pickwell-MacPherson, Y. P. Liang, Y. T. Zhang}},
   doi = {10.1109/iembs.2009.5333091},
   journal = {2009 Annual International Conference of the IEEE Engineering in Medicine and Biology Society},
   month = {9},
   pages = {1746-1749},
   title = {Noninvasive cardiac output estimation using a novel photoplethysmogram index},
   year = {2009},
}

@article{MejiaMejia2022,
   author = {{E. Mejía-Mejía, J. Allen, K. Budidha, C. El-Hajj, P. A. Kyriacou, P. H. Charlton}},
   doi = {10.1016/b978-0-12-823374-0.00015-3},
   journal = {Photoplethysmography},
   month = {11},
   pages = {69-146},
   title = {Photoplethysmography signal processing and synthesis},
   year = {2022},
}

@article{Huang1998,
   author = {{N. E. Huang, Z. Shen, S. R. Long, et al.}},
   doi = {10.1098/rspa.1998.0193},
   journal = {Proceedings of the Royal Society of London. Series A: Mathematical, Physical and Engineering Sciences},
   month = {3},
   number = {1971},
   pages = {903-995},
   title = {The empirical mode decomposition and the Hilbert spectrum for nonlinear and non-stationary time series analysis},
   volume = {454},
   year = {1998},
}

@article{John2022,
   author = {{A. John, S. J. Redmond, B. Cardiff, D. John}},
   doi = {10.1109/jiot.2021.3093112},
   journal = {IEEE Internet of Things Journal},
   month = {2},
   number = {3},
   pages = {2071-2082},
   title = {A Multimodal Data Fusion Technique for Heartbeat Detection in Wearable IoT Sensors},
   volume = {9},
   year = {2022},
}

@article{YZhang2025EMBC,
   author = {{Y. Zhang, L. Fresiello, P. H. Veltink, D. W. Donker, Y. Wang}},
   doi = {10.1109/embc58623.2025.11251595},
   journal = {2025 47th Annual International Conference of the IEEE Engineering in Medicine and Biology Society (EMBC)},
   month = {7},
   pages = {1-5},
   title = {A physiological-model-based neural network framework for blood pressure estimation from photoplethysmography signals},
   year = {2025},
}

@article{G2025,
   author = {{J. G, A. A. Anil, P. M. Nabeel, J. Joseph}},
   doi = {10.1109/EMBC58623.2025.11253561},
   journal = {2025 47th Annual International Conference of the IEEE Engineering in Medicine and Biology Society (EMBC)},
   month = {7},
   pages = {1-5},
   title = {Deep learning-based cardiac output estimation using multimodal physiological signals},
   year = {2025},
}

@misc{Yaowendata2026,
   author = {{Y. Zhang, et al.}},
   doi = {10.5281/zenodo.18935304},
   howpublished = {Zenodo},
   month = {Apr.},
   title = {Finger PPG and Beat-to-beat Blood Pressure in Resting/Cycling},
   year = {2026},
}

@article{Charlton2019,
   author = {{P. H. Charlton, J. Mariscal Harana, S. Vennin, Y. Li, P. Chowienczyk, J. Alastruey}},
   doi = {10.1152/ajpheart.00218.2019},
   journal = {American Journal of Physiology-Heart and Circulatory Physiology},
   month = {11},
   number = {5},
   pages = {H1061-H1085},
   title = {Modelling arterial pulse waves in healthy ageing: a database for in silico evaluation of haemodynamics and pulse wave indices},
   volume = {317},
   year = {2019},
}

@article{Lingawi2025,
   author = {{S. Lingawi, G. Frank, B. H. Kartawidjaja, M. Khalili, B. Kwon, C. Kuo}},
   doi = {10.48550/arXiv.2510.05355},
   journal = {arXiv preprint},
   month = {10},
   title = {Reducing latency and noise in PPG-based SpO2 measurements: a Kalman filtering approach towards acute hypoxia detection},
   year = {2025},
}

@article{Zhang2026,
   author = {{Y. Zhang, L. Fresiello, P. H. Veltink, D. W. Donker, Y. Wang}},
   doi = {10.1016/j.cmpb.2026.109479},
   journal = {Computer Methods and Programs in Biomedicine},
   month = {10},
   pages = {109479},
   title = {{PMB-NN: Physiology-centered hybrid AI for personalized hemodynamic monitoring from photoplethysmography}},
   volume = {285},
   year = {2026},
}

@article{Lakatta2003,
author = {Lakatta, Edward G. and Levy, Daniel},
doi = {10.1161/01.CIR.0000048892.83521.58},
journal = {Circulation},
month = {1},
number = {1},
pages = {139--146},
title = {Arterial and Cardiac Aging: Major Shareholders in Cardiovascular Disease Enterprises},
volume = {107},
year = {2003},
}

@article{Strait2012,
author = {Strait, James B. and Lakatta, Edward G.},
doi = {10.1016/j.hfc.2011.08.011},
journal = {Heart Failure Clinics},
month = {1},
number = {1},
pages = {143--164},
title = {Aging-Associated Cardiovascular Changes and Their Relationship to Heart Failure},
volume = {8},
year = {2012},
}
\end{document}